\documentclass[aps,a4paper,pra,nofootinbib,twocolumn]{revtex4}

\usepackage{amssymb}
\usepackage{amsmath}
\usepackage{amsfonts}
\usepackage{graphicx}
\def\Cas{\mathrm{Cas}}
\def\T{\mathrm{T}}
\def\B{\mathrm{B}}

\def\E{\mathrm{E}}
\def\bk{\mathbf{k}}
\def\br{\mathbf{r}}
\def\dd{\mathrm{d}}
\def\Lif{\mathrm{Lif}}
\def\phit{\widetilde{\phi}}
\def\TE{\mathrm{TE}}
\def\TM{\mathrm{TM}}

\def\P{\mathrm{P}}    
\def\C{\mathrm{C}}    
\def\L{\mathrm{L}}    
\def\R{\mathrm{R}}    
\def\in{\mathrm{in}}  
\def\out{\mathrm{out}}
\def\PP{\mathrm{PP}}  
\def\PS{\mathrm{PS}}  
\def\lat{\mathrm{lat}}
\def\sp{\mathrm{sp}}  
\def\nsp{\mathrm{nsp}}
\def\si{\mathrm{si}}  
\def\GR{G_\mathrm{r}} 
\def\GC{G_\mathrm{c}} 
\def\rR{\rho_\mathrm{r}} 
\def\rC{\rho_\mathrm{c}} 
\def\aR{\alpha_\mathrm{r}} 
\def\aC{\alpha_\mathrm{c}} 

\begin{document}
\title{The Casimir effect within scattering theory}
\author{Astrid Lambrecht$^1$, Paulo A. Maia Neto$^2$, and Serge Reynaud$^1$}
\affiliation{$^1$ Laboratoire Kastler Brossel, CNRS, ENS,
Universit\'e Pierre et Marie Curie case 74, Campus Jussieu, F-75252
Paris Cedex 05, France} \affiliation{$^2$ Instituto de F\'{\i}sica,
UFRJ, CP 68528,   Rio de Janeiro,  RJ, 21941-972, Brazil}

\begin{abstract}
We review the theory of the Casimir effect using scattering
techniques. After years of theoretical efforts, this formalism is
now largely mastered so that the accuracy of theory-experiment
comparisons is determined by the level of precision and pertinence
of the description of experimental conditions. Due to an imperfect
knowledge of the optical properties of real mirrors used in the
experiment, the effect of imperfect reflection remains a source of
uncertainty in theory-experiment comparisons. For the same reason,
the temperature dependence of the Casimir force between dissipative
mirrors remains a matter of debate. We also emphasize that real
mirrors do not obey exactly the assumption of specular reflection,
which is used in nearly all calculations of material and temperature
corrections. This difficulty may be solved by using a more general
scattering formalism accounting for non-specular reflection with
wavevectors and field polarizations mixed. This general formalism
has already been fruitfully used for evaluating the effect of
roughness on the Casimir force as well as the lateral Casimir force
appearing between corrugated surfaces. The commonly used `proximity
force approximation' turns out to lead to inaccuracies in the
description of these two effects.

\end{abstract}
\maketitle

\section{Introduction}

After its prediction in 1948 \cite{Casimir48}, the Casimir force has been observed
in a number of `historic' experiments which confirmed its existence and main
properties \cite{Sparnaay89,Milonni94,Mostepanenko97,LamoreauxResource99}.
With present day technology, a new generation of Casimir force measurements
has started since nearly a decade ago
\cite{Lamoreaux97,Mohideen98,Harris00,Ederth00,Bressi02,Decca03prl,Decca05}.
These experiments have reached a good enough accuracy to allow for
a comparison between theoretical predictions and experimental observations
which is of great interest for various reasons
\cite{Bordag01,Lambrecht02,Milton05}.

The Casimir force is the most accessible effect of vacuum
fluctuations in the macroscopic world. As the existence of vacuum
energy raises difficulties at the interface between the theories of
quantum and gravitational phenomena, it is worth testing this effect
with the greatest care and highest accuracy
\cite{Reynaud01,Genet02Iap}. A precise knowledge of the Casimir
force is also a key point in many accurate force measurements for
distances ranging from nanometer to millimeter. These experiments
are motivated either by tests of Newtonian gravity at millimetric
distances \cite{Fischbach98,Hoyle01,Adelberger02,Long02} or by
searches for new weak forces predicted in theoretical unification
models with nanometric to millimetric ranges
\cite{Carugno97,Bordag99,Fischbach99,Long99,Fischbach01,Decca03prd}.
Basically, they aim at putting limits on deviations of experimental
results from present standard theory. As the Casimir force is the
dominant force between two neutral non-magnetic objects in the range
of interest, any new force would appear as a difference between
experimental measurements and theoretical expectations of the
Casimir force. On a technological side, the Casimir force has been
shown to become important in the architecture of micro- and
nano-oscillators (MEMS, NEMS) \cite{Roukes01,Chan01}. In this
context, it is extremely important to account for the conditions of
real experiments.

\begin{figure}[ptb]
\begin{center} \includegraphics[width=4cm]{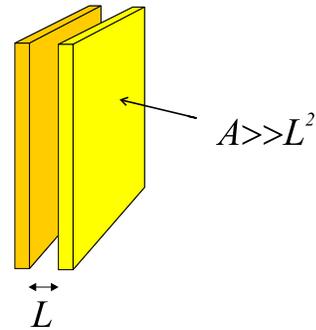}
\caption{Original Casimir configuration of two plane parallel
mirrors a distance $L$ apart.} \label{fig1a}
\end{center}
\end{figure}

The comparison between theory and experiment should take into account the
important differences between the real experimental conditions and the
ideal situation considered by Casimir.
Casimir calculated the force between a pair of perfectly smooth, flat and
parallel plates in the limit of zero temperature and perfect reflection (see Fig.\ref{fig1a}).
He found an expression for the force $F_\Cas$
and the corresponding energy $E_\Cas$ which only depend
on the distance $L$, the area $A$ and two fundamental constants, the speed of light
$c$ and Planck constant $\hbar$
\begin{eqnarray}
F_\Cas &=& \frac{\hbar c \pi ^2 A}{240L^4} =
\frac{\dd E_\Cas}{\dd L} \nonumber \\
E_\Cas &=& - \frac{\hbar c \pi^2 A}{720 L^3} \label{Fcasimir}
\end{eqnarray}
Each transverse dimension of the plates has been supposed to be much
larger than $L$. Conventions of sign have been chosen so that
$F_\Cas$ is positive while $E_\Cas$ is negative. They correspond to
an attractive force ($\sim 0.1 \mu$N for $A=1 {\rm cm}^2$ and
$L=1\mu$m) and a binding energy.

The fact that the Casimir force (\ref{Fcasimir}) only depends on fundamental
constants and geometrical features is remarkable. In particular it is independent
of the fine structure constant which appears in the expression of the atomic
Van der Waals forces. This universality property is related to the assumption
of perfect reflection used by Casimir in his derivation. Perfect mirrors
correspond to a saturated response to the fields since they reflect 100\%
of the incoming light. This explains why the Casimir effect, though it has its
microscopic origin in the interaction of electrons with electromagnetic fields,
does not depend on the fine structure constant.

However, no real mirror can be considered as a perfect reflector at all field
frequencies. In particular, the most precise experiments are performed with
metallic mirrors which show perfect reflection only at frequencies smaller
than a characteristic plasma frequency $\omega_\P$ which depends on the
properties of conduction electrons in the metal. Hence the Casimir force
between metal plates can fit the ideal Casimir formula (\ref{Fcasimir}) only
at distances $L$ much larger than the plasma wavelength
\begin{equation}
\lambda _\P=\frac{2\pi c}{\omega _\P}
\end{equation}
For metals used in the recent experiments, this wavelength lies in the
$0.1\mu$m range (107nm for Al and 137nm for Cu and Au).
At distances smaller than or of the order of the plasma wavelength, the finite
conductivity of the metal has a significant effect on the force.
The idea has been known since a long time \cite{Lifshitz56,Heinrichs75,Schwinger78}
but a precise quantitative investigation of the effect of imperfect reflection
has been systematically developed only recently
\cite{Lamoreaux99,Lambrecht00,KlimPRA00,MostepanenkoPRA00}.
As the effect of imperfect reflection is large in the most accurate experiments,
a precise knowledge of its frequency dependence is essential for obtaining an
accurate theoretical prediction of the Casimir force.

This is also true for other corrections to the ideal Casimir formula
associated with the experimental configuration. For experiments at
room temperature, the effect of thermal field fluctuations,
superimposed to that of vacuum, affects the Casimir force at
distances larger than a few microns. Again the idea has been known
for a long time \cite{Mehra67,Brown69} but a quantitative evaluation
taking into account the correlation of this effect with that of
imperfect reflection has been mastered only recently
\cite{Genet00,Reynaud03}. A number of publications have given rise
to contradictory estimations of the Casimir force between
dissipative mirrors at non zero temperature
\cite{Bostrom00,Svetovoy00,Bordag00,Klimchitskaya01,Hoye03}. Many
attempts have been made to elucidate the problem by taking into
account the low-frequency character of the force between metallic
films \cite{Lamoreaux04}, the spatial dispersion on electromagnetic
surface modes \cite{Sernelius05} or the transverse momentum
dependance of surface impedances
\cite{Svetovoy04,BrevikPRE05,Hoye05}. Experimentally the effect of
temperature of the Casimir force has not yet be conclusively
measured \cite{Mostepanenko05}. For a recent analysis of this issue
see reference \cite{BrevikNJP06}.

Most experiments are performed between a plane and a sphere with the
force estimation involving a geometry correction. Usually the
Casimir force in the plane-sphere (PS) geometry is calculated using
the Proximity Force Approximation (PFA). This approximation amounts
to the addition of force or energy contributions corresponding to
different local inter-plate distances, assuming these contributions
to be independent. But the Casimir force and energy are not
additive, so that the PFA cannot be exact, although it is often
improperly called a theorem.

In the present review, we consider both the original Casimir
geometry with perfectly plane and parallel mirrors and the
plane-sphere geometry when comparing to experiments. The PFA is
expected to be valid in the plane-sphere geometry, when the sphere
radius $R$ is much larger than the separation $L$
\cite{Derjaguin68,Langbein71,Kiefer78}, which is the case for all
present day experiments, and it will thus be used to connect the two
geometries. In this case, the force $F_\PS$ between a sphere of
radius $R$ and a plane at a distance of closest approach $L$ is
given in terms of the energy $E_\PP$ for the plane-plane cavity as
follows
\begin{equation}
F_\PS  =2\pi R\frac{E_\PP}{A} \quad,\quad L\ll R
\label{FPSsmooth}
\end{equation}
Interesting attempts to go beyond this approximation concerning the plane-sphere geometry
have been made recently \cite{JaffePRA05,Gies06}, in the more general context of the connection
between geometry and the Casimir effect \cite{Balian78,Plunien86,Balian04}.

Another important correction to the Casimir force is coming from
surface roughness, which is intrinsic to any real mirror, with
amplitude and spectrum varying depending on the surface preparation
techniques. The departure from flatness of the metallic plates may
also be designed, in particular under the form of sinusoidal
corrugation of the plates which produce a measurable lateral
component of the Casimir force \cite{ChenPRL02}. For a long time,
these roughness or corrugation corrections to the Casimir force have
been calculated with methods valid only in limiting cases
\cite{Bree74,Maradudin75,Maradudin80,Vesperinas82,EmigPRL01,EmigEPL03}
or by using PFA \cite{BordagPLA95,KlimPRA99,BlagovPRA04}. Once
again, it is only recently that emphasis has been put on the
necessity of a more general method for evaluating the effect of
roughness outside the region of validity of PFA with imperfect
mirrors at arbitrary distances from each other \cite{GenetEPL03}.
While the condition $L\ll R$ is sufficient for applying the PFA in
the plane-sphere geometry, more stringent conditions are needed for
PFA to hold for rough or corrugated surfaces. The surfaces should
indeed be nearly plane when looked at on a scale comparable with the
separation $L$, and this condition is not always satisfied in
experiments. When PFA is no longer valid, the effect of roughness or
corrugation can be evaluated by using the scattering theory extended
to the case of non-specular reflection.

We review the Casimir effect within scattering theory and the theory of quantum optical networks.
The main idea of this derivation is that the Casimir force has its origin in a difference of the
radiation pressure of vacuum fields between the two mirrors and in the outer free field vacuum.
This vacuum radiation pressure can be written as an integral over all modes,
each mode being associated with reflection amplitudes on the two mirrors.
We first present formulas written for specular reflection which are valid for
lossless\cite{Jaekel91} as well as lossy mirrors \cite{GenetPRA03}.
We discuss the influence of the mirrors reflection coefficients at zero and non-zero temperature.
We then extend the approach to the case of non-specular reflection which mixes the
field polarizations and transverse wave-vectors.
Finally we apply the latter approach to the calculation of the roughness correction
to the Casimir force between metallic mirrors \cite{MaiaEPL05,MaiaPRA05} and of the lateral
component of the Casimir force between corrugated plates \cite{MaiaPRL06}.

\section{Specular scattering}

\begin{figure}[ptb]
\begin{center} \includegraphics[width=7cm]{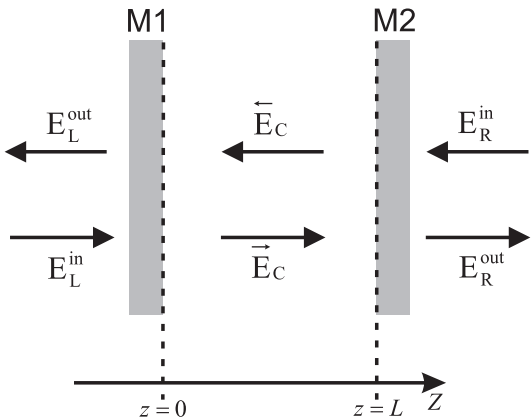}
\caption{Schematic view of a Fabry-Perot cavity of length $L$.}
\label{fig1b}
\end{center}
\end{figure}

Let us first consider the original Casimir geometry with perfectly plane and parallel mirrors
aligned along the directions $x$ and $y$.
The two mirrors thus form a Fabry-Perot cavity of length $L$  as shown in Fig.\ref{fig1b}.
We analyze the cavity as a composed optical network, and calculate the fluctuations
of the intracavity fields propagating along the positive and negative $z$-axis,
$\stackrel{\rightarrow}{\E}_\C$ and $\stackrel{\leftarrow}{\E}_\C$,
in terms of the fluctuations of the incoming free-space fields $\E_\L^\in$ and $\E_\R^\in$
(the outgoing fields $\E_\L^\out$ and $\E_\R^\out$ are also shown).

The field modes are conveniently characterized by their frequency $\omega$,
transverse wavevector $\bk$ with components $k_{x},k_{y}$ in the plane of the mirrors
and polarization $p$.
As the configuration of Fig.\ref{fig1b} obeys a symmetry with respect to time translation
as well as transverse space translations (along directions $x$ and $y$),
the frequency $\omega $, transverse vector $\bk \equiv \left( k_{x},k_{y}\right)$ and
polarization $p=\TE,\TM$ are preserved throughout the
whole scattering processes on a mirror or a cavity. The scattering couples
only the free vacuum modes which have the same values for the preserved quantum
 numbers and differ by the sign of the longitudinal component $k_{z}$ of the wavevector.
We denote by $\left(r_\bk^p\left[\omega\right]\right)_j$ the reflection amplitude
of the mirror $j=1,2$ as seen from the inner side of the cavity.
This scattering amplitude obeys general properties of causality,
unitarity and high frequency transparency. The additional
fluctuations accompanying losses inside the mirrors are deduced from
the optical theorem applied to the scattering process which couples
the modes of interest and the noise modes \cite{Barnett96,Courty00}.

The loop functions which characterize the optical response of the
cavity to an input field play an important role in the following
\begin{eqnarray}
f_\bk^p \left[\omega\right] &=&
\frac{\rho_\bk^p\left[\omega\right]}{1-\rho_\bk^p\left[\omega\right]}\nonumber\\
\rho_\bk^p\left[\omega\right] &=&
\left(r_\bk^p\left[\omega\right]\right)_1
\left(r_\bk^p\left[\omega\right]\right)_2 e^{2ik_z L}\label{rho}
\end{eqnarray}
$\rho_\bk^p$ and $f_\bk^p$ are respectively the open-loop and
closed-loop functions corresponding to one round trip in the cavity.
The system formed by the mirrors and fields is stable so that
$f_\bk^p$ is an analytic function of frequency $\omega$. Analyticity
is defined with the following physical conditions in the complex
plane
\begin{eqnarray}
\omega\equiv i\xi &&\quad,\quad \Re\xi > 0 \label{PhysConditions} \\
k_z\equiv i\kappa\left[\omega\right] \quad,\quad&&
\kappa\left[\omega\right]
\equiv\sqrt{\bk^2-\frac{\omega^2}{c^2}}
\quad,\quad\Re\kappa\left[\omega\right] > 0 \nonumber
\end{eqnarray}
The quantum numbers $p$ and $\bk$ remain spectator throughout the
discussion of analyticity. The sum on transverse wavevectors may be
represented as a sum over the eigenvectors
$k_x=2\pi q_x/L_x,k_y=2\pi q_y/L_y$ associated with virtual quantization boxes
along $x,y$ or, at the continuum limit $L_x,L_y\to\infty$ with $A=L_xL_y$,
as an integral
\begin{eqnarray}
\sum_\bk &\equiv& \sum_{q_x=-\infty}^{\infty}
\sum_{q_y=-\infty}^{\infty} \to A \int_{-\infty}^{\infty} \frac{\dd
k_x}{2\pi} \int_{-\infty}^{\infty} \frac{\dd k_y}{2\pi}
\end{eqnarray}

We then introduce the Airy function defined in classical optics as
the ratio of energy inside the cavity to energy outside the cavity
for a given mode
\begin{eqnarray}
&&g_\bk^p\left[\omega\right] = 1 + \left\{
f_\bk^p\left[\omega\right] + c.c. \right\} =
\frac{1-\left|\rho_\bk^p\left[\omega\right]\right| ^2} {\left|
1-\rho_\bk^p\left[\omega\right] \right| ^2}
\label{Airyfunction}
\end{eqnarray}
$f_\bk^p$, $g_\bk^p$ depend only on the reflection amplitudes of
mirrors as they are seen from the inner side. With these definitions
we write the Casimir force
\begin{eqnarray}
F &=&  -\hbar \sum_p \sum_\bk \int_{0}^{\infty}
\frac{\dd\omega}{2\pi} \left\{ i \kappa\left[\omega\right]
f_\bk^p\left[\omega\right]+ c.c. \right\}
\label{ForceReal}
\end{eqnarray}
or, equivalently, the Casimir energy
\begin{eqnarray}
E &=&  -\hbar \sum_p \sum_\bk \int_{0}^{\infty}
\frac{\dd\omega}{2\pi} \frac{1}{2i}
\ln \left[ \frac{1-\rho_\bk^p[\omega]}{1-\rho_\bk^p[\omega]^*}\right]
\label{EReal}
\end{eqnarray}
Equations (\ref{ForceReal},\ref{EReal}) contain the contribution of ordinary
modes freely propagating outside and inside the cavity with $\omega > c |\bk|$
and $k_z$ real. This contribution thus merely reflects
the intuitive picture of a radiation pressure of fluctuations on the
mirrors of the cavity \cite{Jaekel91} with the factor $g_\bk^p-1$
representing a difference between inner and outer sides. Equations
(\ref{ForceReal},\ref{EReal}) also include the contribution of evanescent waves
with $\omega < c |\bk|$ and $k_z$ imaginary. Those waves propagate
inside the mirrors with an incidence angle larger than the limit
angle and they also exert a radiation pressure on the mirrors, due
to the frustrated reflection phenomenon \cite{GenetPRA03}. Their
properties are conveniently described through an analytical
continuation of those of ordinary waves, using the well defined
analytic behavior of $\kappa$ and $f_\bk^p$.

Using analyticity properties, we now transform
(\ref{ForceReal}) into an integral over imaginary frequencies by
applying the Cauchy theorem on the contour enclosing the quadrant
$\Re\omega>0,\Im\omega>0$. We use high frequency transparency to
neglect the contribution of large frequencies. This leads to the
following expression for the Casimir force
\begin{eqnarray}
F = \hbar \sum_p\sum_\bk\int_{0}^{\infty}\frac{\dd\xi}{2\pi} \left\{
\kappa\left[i\xi\right] f_\bk^p \left[i\xi\right] +c.c. \right\}
\label{ForceImag}
\end{eqnarray}
which is now written as an integral over complex frequencies $\omega=i\xi$.
In the same way we obtain the Casimir energy as a function of imaginary frequencies
\begin{eqnarray}
E=\frac{\hbar A}{2\pi}\sum_p \int \frac{\dd^2\bk}{4\pi ^2}
\int\limits_{0}^{\infty} \dd\xi \ \ln\left[1-\rho_\bk^p \left[ i\xi
\right] \right] \label{Eim}
\end{eqnarray}
Causality and passivity conditions assure that the integrand
$\ln\left[1-\rho_\bk^p \left[ i\xi \right] \right]$
is analytical in the upper half space of the complex plane $\Re\xi>0$.
It is thus clear that both expressions for force and energy are equivalent.

\subsection{Finite conductivity correction}

Let us now review the correction to the Casimir force coming from
the finite conductivity of any material. This correction is given by
relations (\ref{ForceReal}) or, equivalently, (\ref{ForceImag}), as
soon as the reflection amplitudes are known. These amplitudes are
commonly deduced from models of mirrors, in particular bulk mirrors,
slabs or layered mirrors, the optical response of metallic matter
being described by some permittivity function. This function may be
either a simple description of conduction electrons in terms of a
plasma or Drude model or a more elaborate representation based upon
tabulated optical data. At the end of the section, we will discuss
the uncertainty in the theoretical evaluation of the Casimir force
coming from the lack of knowledge of the specific material
properties of a given mirror as was illustrated in
\cite{Lambrecht00}.

Assuming that the metal plates have a large optical thickness,
the reflection coefficients correspond to the ones of a simple
vacuum-bulk interface \cite{LandauECM9}
\begin{eqnarray}
r^\TE &=&-\frac{\sqrt{\xi ^{2}\left( \varepsilon \left( i\xi
\right) -1\right) +c^{2}\kappa ^{2}}-c\kappa }{\sqrt{\xi ^{2}\left(
\varepsilon \left( i\xi \right) -1\right) +c^{2}\kappa ^{2}}+c\kappa }
\nonumber \\
r^\TM &=&\frac{\sqrt{\xi ^{2}\left( \varepsilon \left( i\xi \right)
-1\right) +c^{2}\kappa ^{2}}-c\kappa \varepsilon \left( i\xi \right) }
{\sqrt{\xi ^{2}\left( \varepsilon \left( i\xi \right) -1\right)
+c^{2}\kappa ^{2}}+c\kappa \varepsilon \left( i\xi \right) }
\label{rThick}
\end{eqnarray}
$r^p$ stands for $r^p \left( i\xi ,i\kappa \right) $ and
$\varepsilon \left( i\xi\right) $ is the dielectric function of the metal
evaluated for imaginary frequencies; the index $\bk$ has been dropped.

Taken together, the relations (\ref{ForceImag},\ref{rThick})
reproduce the Lifshitz expression for the Casimir force
\cite{Lifshitz56}. Note that the expression was not written in this
manner by Lifshitz. To our present knowledge, Kats \cite{Kats77} was
the first to stress that Lifshitz expression could be written in
terms of the reflection amplitudes (\ref{rThick}). We then have to
emphasize that (\ref{ForceImag}) is much more general than Lifshitz
expression since it still holds with mirrors characterized by
reflection amplitudes differing from (\ref{rThick}). As an
illustration, we may consider metallic slabs having a finite
thickness.

For a given polarization, we denote by $r_\si$ the reflection
coefficient (\ref{rThick}) corresponding to a single vacuum/metal interface
and we write the reflection amplitude $r$ for the slab of finite thickness
through a Fabry-P\'{e}rot formula
\begin{eqnarray}
r &=&r_\si \frac{1-e^{-2\delta }}{1-r_\si^2 e^{-2\delta }}  \nonumber \\
\delta &=&\frac{D}{c}\sqrt{\xi ^{2}\left( \varepsilon \left( i\xi
\right) -1\right) +c^{2}\kappa ^{2}}  \label{rSlab}
\end{eqnarray}
This expression has been written directly for imaginary frequencies. The
parameter $\delta$ represents the optical length in the metallic slab and
$D$ the physical thickness. The single interface expression (\ref{rThick})
is recovered in the limit of a large optical thickness $\delta \gg 1$. With
the plasma model, this condition just means that the thickness $D$ is larger
than the plasma wavelength $\lambda _\P $.

\begin{figure}[tbh]
\begin{center} \includegraphics[width=3cm]{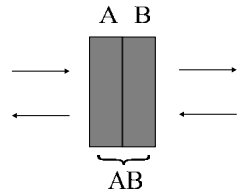}
\caption{\label{multi}Composition of networks~: two networks labeled
A and B are piled up to build up a network AB.}
\end{center}
\end{figure}

In order to discuss experiments, it may also be worth to write the
reflection coefficients for multilayer mirrors. For example one may consider
two-layer mirrors with a layer of thickness $D$ of a metal $A$ deposited on
a large slab of metal $B$ in the limit of large thickness as shown in Figure \ref{multi}.
The reflection formulas are then obtained as in \cite{Lambrecht97} but accounting for
oblique incidence
\begin{eqnarray}
r_{AB} =r_{A}+\frac{t_{A}^{2} r_{B}}{1-r_{A} r_{B}}  \quad,\quad
t_{AB} =\frac{t_{A} t_{B}}{1-r_{A} r_{B}}
\label{compoSlab}
\end{eqnarray}
It reproduces the known results for the simple multilayer systems which have
already been studied \cite{Bordag01}.
The combination of (\ref{rSlab}) and (\ref{compoSlab}) allows to calculate most
of the experimental situations precisely.

In order to assess quantitatively the effect of finite conductivity,
we may in a first approach use the plasma model for the metallic
dielectric function, with $\omega _\P$ the plasma frequency,
\begin{eqnarray}
\varepsilon \left( \omega \right) =1-\frac{\omega _\P ^{2}}{\omega ^{2}} \quad,\quad
\varepsilon \left( i\xi \right) =1+\frac{\omega _\P ^{2}}{\xi ^{2}}
\label{epsPlasma}
\end{eqnarray}
It is convenient to present the change in the Casimir force in terms of a
factor $\eta _{F}$ which measures the reduction of the force with respect to
the case of perfect mirrors
\begin{equation}
F=\eta _{F}F_{C}  \label{defetaF}
\end{equation}
Using expressions (\ref{rThick},\ref{epsPlasma}) it is possible to obtain
the reduction factor defined for the Casimir force through
numerical integrations.

\begin{figure}[tbp]
\begin{center} \includegraphics[width=7cm]{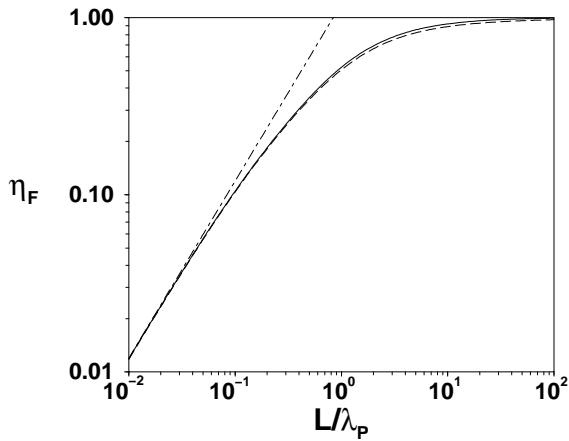}
\caption{Reduction of the Casimir force compared to the force
between perfect mirrors, when the finite conductivity is described
by a plasma model (solid line) or a Drude model (dashed line) with a
ratio $\frac{\gamma }{\omega _\P }$ equal to $4 \times 10^{-3}$. The
difference due to the relaxation parameter has only a small effect
on the calculation of the Casimir force. The dotted-dashed line
corresponds to the short distance asymptotic behavior
(\ref{etashortF}).} \label{fig-plasma}
\end{center}
\end{figure}

The result is plotted as the solid line on figure \ref{fig-plasma}, as a
function of the dimensionless parameter $\frac{L}{\lambda _\P }$, that is
the ratio between the distance $L$ and the plasma wavelength $\lambda _\P $.
As expected the Casimir formula is reproduced at large distances
($\eta_{F}\rightarrow 1$ when $L\gg\lambda_\P$). At distances smaller
than $\lambda _\P $ in contrast, a significant reduction is obtained with
the asymptotic law of variation read as \cite{GenetAFLdB04,Henkel04}
\begin{eqnarray}
L &\ll &\lambda _\P \quad \rightarrow \quad \eta _{F}=\alpha \frac{L}{%
\lambda _\P }  \quad \alpha \simeq 1.193  \label{etashortF}
\end{eqnarray}
This can be understood as the result of the Coulomb interaction of
surface plasmons at the two vacuum/metal interfaces
\cite{Kampen68,GenetAFLdB04}. The generalization of this idea at
arbitrary distances is more subtle since it involves a full
electromagnetic treatment of the plasmon as well as ordinary photon
modes \cite{plasmon}.

The plasma model cannot provide a fully satisfactory description of the
optical response of metals, in particular because it does not account
for any dissipative mechanism.
A more realistic representation is the Drude model \cite{AshcroftMermin}
\begin{eqnarray}
\varepsilon \left( \omega \right) &=&1-\frac{\omega _\P ^2}{\omega \left(
\omega +i\gamma \right) }  \nonumber \\
\varepsilon \left( i\xi \right) &=&1+\frac{\omega _\P ^2}{\xi \left(
\xi +\gamma \right) }
\label{epsDrude}
\end{eqnarray}
This model describes not only the plasma response of conduction electrons
with $\omega _\P $ still interpreted as the plasma frequency but also their
relaxation, $\gamma $ being the inverse of the electronic relaxation time.

The relaxation parameter $\gamma $ is much smaller than the plasma
frequency. For Al, Au, Cu in particular, the ratio $\gamma /\omega
_\P$ is of the order of $4\times 10^{-3}$. Hence relaxation affects
the dielectric constant in a significant manner only at frequencies
where the latter is much larger than unity. In this region, the
metallic mirrors behave as a nearly perfect reflectors so that,
finally, the relaxation does not have a large influence on the
Casimir effect at zero temperature. This qualitative discussion is
confirmed by the result of numerical integration reported as the
dashed line on figure \ref{fig-plasma}. With the typical value
already given for $\gamma /\omega _\P$, the variation of $\eta _{F}$
remains everywhere smaller than 2\%.

For metals like Al, Au, Cu, the dielectric constant departs from the
Drude model when interband transitions are reached, that is when the
photon energy reaches a few eV. Hence, a more precise description of
the dielectric constant should be used for evaluating the Casimir
force in the sub-$\mu$m range. This description relies on one hand
on the causality relations obeyed by the dielectric response
function and on another hand on known optical data. The reader is
referred to \cite{Lambrecht00} for a detailed analysis, but we
recall here the main argument and some important details. Let us
first recall that frequencies are measured either in $e$V or in
rad/s, using the equivalence 1~eV $=1.519\times 10^{15}$~rad/s. An
erroneous conversion factor 1~eV $=1.537\times 10^{15}$~rad/s was
used in \cite{Lambrecht00}, which led to a difference in
$\varepsilon(i\xi)$ of less than 1\% over the relevant distance
range. In the end of the calculation, this was corresponding to a
negligible error in the Casimir force and energy \cite{data}.

The values of the complex index of refraction for different metals, measured through different
optical techniques, are tabulated as a function of frequency in several
handbooks \cite{Palik,McGrawHill,CRC98}. Optical data may vary from one
reference to another, not only because of experimental uncertainties but
also because of the dispersion of material properties of the analyzed samples.
Moreover, the available data do not cover a broad enough frequency range
so that they have to be extrapolated.
These problems may cause variations of the results obtained for the dielectric
function $\varepsilon \left( i\xi\right) $ and, therefore, for the Casimir force.

\begin{figure}[tbh]
\begin{center} \includegraphics[width=7cm]{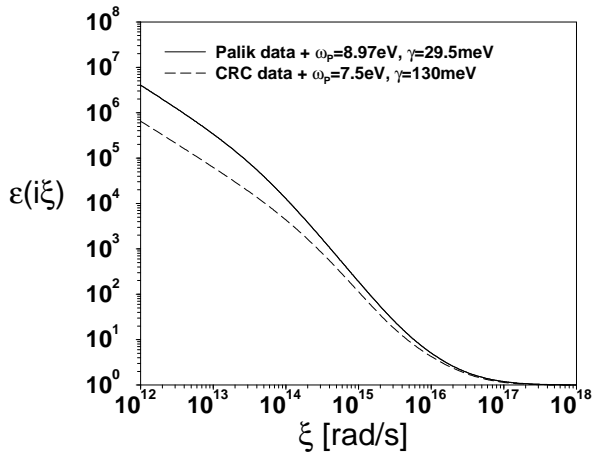}
\caption{Dielectric function of Cu versus imaginary frequency. The
solid line corresponds to the first set (optical data given by
\protect\cite{Palik,McGrawHill}, low frequencies extrapolated by a
Drude model with $\omega _\P =8.97$~eV and $\gamma =29.5$~meV), the
dashed line to the second set (optical data given by
\protect\cite{CRC98}, low frequencies extrapolated by a Drude model
with $\omega _\P =7.5$~eV and $\gamma =130$~me V).}
\label{fig-eiwCu}
\end{center}
\end{figure}

Figure \ref{fig-eiwCu} shows two different plots of $\varepsilon \left( i\xi\right) $
for Cu as a function of imaginary frequency $\xi$.
The solid line corresponds to the first data set  with data points taken from
\cite{Palik,McGrawHill} and extrapolation at low frequency with a Drude model
with parameters $\omega _\P =8.97$~eV and $\gamma =29.5$~meV
in reasonable agreement with existing knowledge from solid state physics.
However, as explained in \cite{Lambrecht00}, the optical data available for
Cu do not permit an unambiguous estimation of the two parameters $\omega_\P$
and $\gamma$ separately. Other couples of values can be chosen which
are also consistent with optical data. To make this point explicit, we
have drawn a second plot on figure \ref{fig-eiwCu} (dashed line) with data
taken from \cite{CRC98} and the low frequency interpolation given by a Drude
model with $\omega _\P =7.5$~eV and $\gamma =130$~meV. These values lead to a
dielectric function $\varepsilon(i\xi)$ smaller than in the first data set over
the whole frequency range, but especially at low frequencies.
An estimation of the uncertainties associated with this imperfect knowledge
of optical data can be drawn from the computation of the Casimir force in
these two cases.

\begin{figure}[tbh]
\begin{center} \includegraphics[width=7cm]{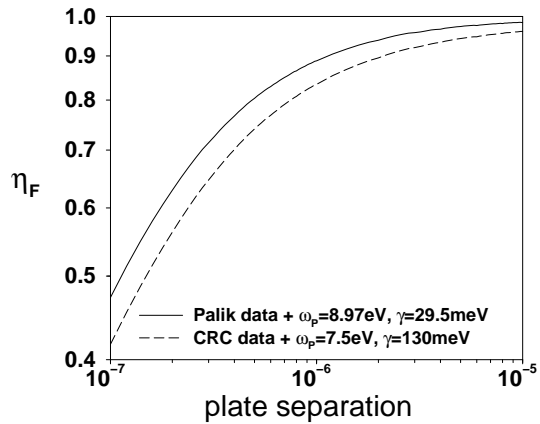}
\caption{Reduction factor $\eta_F$ for the Casimir force between two
Cu plates as a function of the plate separation $L$. The solid line
corresponds to the first set (optical data given by
\protect\cite{Palik}, low frequencies extrapolated by a Drude model
with $\omega _\P =8.97$~eV and $\gamma =29.5$~meV), the dashed line
to the second set (optical data given by \protect\cite{CRC98}, low
frequencies extrapolated by a Drude model with $\omega _\P =7.5$~eV
and $\gamma =130$~m$e$V).} \label{fig-etaCu}
\end{center}
\end{figure}

Figure \ref{fig-etaCu} shows the reduction factor $\eta_F$ for the
Casimir force between two Cu plates as a function of the plate
separation $L$  for the two sets of optical data. The two
corresponding curves have similar dependance on the plate separation
but the absolute values are shifted from one curve to the other. At
a separation of 100 nm the difference can be as large as 5\%. As the
plasma frequency is basically the frequency above which the mirrors
reflectivity diminishes considerably, the Drude parameters of the
first set ($\omega_\P =8.97$ eV and $\gamma=29.5$ meV) give a larger
Casimir force than the second set, where the plasma frequency is
lower ($\omega_\P =7.5$ eV and $\gamma=130$ meV). A detailed
analysis of this uncertainty has been recently reported
\cite{Pirozhenko06}.

Let us emphasize that the problem here is neither due to a lack of precision
of the calculations nor to inaccuracies in experiments.
The problem is that calculations and experiments may consider
physical systems with different optical properties.
Material properties of mirrors indeed vary considerably as a function of external
parameters and preparation procedure \cite{Pirozhenko06}.
This difficulty could be solved by measuring the reflection amplitudes
of the mirrors used in the experiment and then inserting these informations
in the formula giving the predicted Casimir force.
In order to suppress the uncertainty associated with the extrapolation procedure,
it would be necessary to measure the reflection amplitudes down to frequencies
of the order of 1 meV, if the aim is to calculate the Casimir force in the
distance range from 100 nm to a few $\mu$m.

\subsection{Temperature correction}

The Casimir force between dissipative metallic mirrors at non zero temperature
has given rise to contradictory claims which have raised doubts about
the theoretical expression of the force. In order to contribute to the resolution
of this difficulty, we now review briefly the derivation of the force from basic
principles of the quantum theory of lossy optical cavities at non zero temperature.
We obtain an expression which is valid for arbitrary mirrors, including
dissipative ones, characterized by frequency dependent reflection amplitudes.
This expressions coincides with the usual Lifshitz expression when the plasma model
is used to describe the mirrors material properties, but it differs when the Drude
model is applied. The difference can be traced back to the validity of Poisson
summation formula \cite{Reynaud03}.

To discuss the effect of finite temperature we use a theorem which
gives the commutators of the intracavity fields as the product of
those well known for fields outside the cavity by the Airy function.
This theorem was demonstrated with an increasing range of validity
in \cite{Jaekel91}, \cite{Barnett98} and \cite{GenetPRA03}. It is
true regardless of whether the mirrors are lossy or not. Since it
does not depend on the state of the field, it can be used for
thermal as well as vacuum fluctuations. Assuming thermal
equilibrium, the theorem leads to the expression of the field
anticommutators, \textit{i.e.} the field fluctuations. Note that
thermal equilibrium has to be assumed for the whole system, which
means that input fields as well as fluctuations associated with
electrons, phonons and any loss mechanism inside the mirrors
correspond to the same temperature $T$, whatever their microscopic
origin may be. If parts of the system correspond to different
temperatures, completely different results are obtained
\cite{Antezza05,Henkel02}.

The anticommutators of intracavity fields are given by those known
for fields outside the cavity multiplied by the Airy function.
Hence, the expression written in \cite{GenetPRA03} for a null
temperature is only modified through the appearance of a thermal
factor in the integrand
\begin{eqnarray}
F &=&  -\hbar \sum_{p,\bk} \int_{0}^{\infty}
\frac{\dd\omega}{2\pi} \left\{ i \kappa_\bk\left[\omega\right]
f_\bk^p\left[\omega\right]
c\left[\omega\right] + c.c. \right\} \nonumber \\
&&c\left[\omega\right] \equiv \coth\left(
\frac{\pi\omega}{\omega_\T} \right) \quad,\quad \omega_\T \equiv
\frac{2\pi k_\B T}{\hbar}  \label{ForceRealT}
\end{eqnarray}

Using as before analyticity properties, we transform
(\ref{ForceRealT}) into an integral over imaginary frequencies giving
the following expression for the Casimir force
\begin{eqnarray}
F = \hbar \sum_{p,\bk}\int_{0}^{\infty}\frac{\dd\xi}{2\pi} \left\{
\kappa\left[i\xi+\eta\right] f_\bk^p \left[i\xi+\eta\right]
c\left[i\xi+\eta\right] +c.c. \right\} && \label{ForceImagT}
\end{eqnarray}
It is now written as an integral over complex frequencies
$\omega=i\xi +\eta$ close to the imaginary axis, with the small
positive real number $\eta \to 0^+$ maintaining the Matsubara poles
$\omega_m=im\omega_\T$ of $c\left[\omega\right]$ outside the contour
used to apply the Cauchy theorem. Up to this point, the present
derivation is similar to Lifshitz' demonstration \cite{Lifshitz56}
while being valid for arbitrary reflection amplitudes. The next
steps in Lifshitz' derivation, scrutinized in \cite{Reynaud03},
may raise difficulties for dissipative mirrors.
Let us briefly recall the main arguments of \cite{Reynaud03}.

We may first write a series expansion of the Casimir force
(\ref{ForceRealT}) based upon the expansion of the function
$\coth\left(\frac{\pi\omega}{\omega_\T} \right)$
into a series of exponentials
$\exp\left(-\frac{2n\pi\imath\xi}{\omega_\T}\right)$
(see also \cite{Genet00}).
This expansion obeys the mathematical criterion of uniform convergence
so that, when it is inserted in (\ref{ForceRealT}), the order of the summation over $n$
and integration over $\xi$ may be exchanged. It follows that the
force (\ref{ForceRealT}) may also be read as
\begin{eqnarray}
F&=& \frac\hbar\pi \sum_p\sum_\bk \sum_n^\prime
\phit_\bk^p \left( \frac{2n\pi}{\omega_\T}\right) \label{ForceImagExp} \\
\phit_\bk^p \left(x\right) &\equiv& 2 \int_0^\infty \dd\xi
\cos\left(\xi x\right) \phi_\bk^p \left[\xi\right] \nonumber\\
\phi_\bk^p \left[\xi\right] &\equiv& \lim_{\eta\to 0^+}
\kappa_\bk\left[i\xi+\eta\right] f_\bk^p \left[i\xi+\eta\right]
\nonumber
\end{eqnarray}
We have introduced the common summation convention
\begin{eqnarray}
&&  \sum_n^\prime \varphi\left(n\right) \equiv
\frac{1}{2}\varphi\left(0\right)
+\sum_{n=1}^{\infty}\varphi\left(n\right) \label{SumConv}
\end{eqnarray}
The function $\phi_\bk^p$ is well defined almost everywhere, the
only possible exception being the point $\xi=0$ where the limit
$\eta \to 0^+$ may be ill defined for mirrors described by
dissipative optical models \cite{Klimchitskaya01}. Since this is a
domain of null measure, the cosine Fourier transform $\phit_\bk^p$
of $\phi_\bk^p$ is well defined everywhere and the expression
(\ref{ForceImagExp}) of the Casimir force is valid for arbitrary
mirrors, including dissipative ones. Note that the term $n=0$ in
(\ref{ForceImagExp}) corresponds exactly to the contribution of
vacuum fluctuations, or to the zero temperature limit, while the
terms $n\geq 1$ give the corrections associated with thermal fields.

We come back to the derivation of the Lifshitz formula
\cite{Lifshitz56}, often used as the standard expression of the
Casimir force. This formula is directly related to the decomposition
of the $\coth$ function into elementary fractions corresponding to
the Matsubara poles $\Omega_m=im\omega_\T$. If we assume furthermore
that the function $\phi_\bk^p$ is a sufficiently smooth test
function, in the sense defined by the theory of distributions, we
deduce that the expression (\ref{ForceImagT}) can also be read
\begin{equation}
F_\Lif = \frac{\hbar\omega_\T}\pi \sum_p\sum_\bk \sum_m^\prime
\phi_\bk^p \left[m\omega_\T\right] \label{LifshitzForm}
\end{equation}
This is the generalization of the Lifshitz' formula
\cite{Lifshitz56} to the case of arbitrary reflection amplitudes. It
is a discrete sum over Matsubara poles with the primed summation
symbol having the definition (\ref{SumConv}). This formula is known
to lead to the correct result in the case of dielectric mirrors (for
which it was derived in \cite{Lifshitz56}), for perfect mirrors
\cite{Mehra67,Brown69} and also for metallic mirrors described by
the lossless plasma model \cite{Genet00}.

However its applicability to arbitrary mirrors remains a matter of
controversy \cite{Klimchitskaya01}.
The point is that the derivation of the Lifshitz' formula
(\ref{LifshitzForm}) requires that the function $\phi_\bk^p$ be a sufficiently
smooth test function, in the sense defined by the theory of distributions.
Whether or not this is the case at $\xi=0$ for $\phi_\bk^p$ calculated from
dissipative optical models constitutes the central question of the
controversy on the value of the term $p$=TE, $m=0$ in Lifshitz' sum
\cite{Bostrom00,Svetovoy00,Bordag00,Klimchitskaya01,Hoye03}. Let us
repeat that (\ref{ForceImagExp}) is still a mathematically valid
expression of the Casimir force even when $\phi_\bk^p$ is ill
defined for $\xi$ in a domain of null measure.
The question of validity of Lifshitz' formula (\ref{LifshitzForm})
may also be phrased in terms of applicability of the Poisson
summation formula \cite{MorseFeshbach}.
This applicability depends on a smoothness condition which is met
for dielectric mirrors, for perfect mirrors and for mirrors described
by the plasma model and this explains why Lifshitz' formula
(\ref{LifshitzForm}) may be used as well as (\ref{ForceImagExp})
in these cases \cite{Reynaud03}.

In order to solve this controversy, it is crucial to improve our
knowledge of the reflection amplitudes at low frequencies. As
already discussed, the best manner to do that is to measure these
amplitudes on the mirrors used in the experiment at frequencies as
low as possible. Although the theoretical question of a good
modeling of mirrors at low frequencies is certainly of interest and
needs to be answered, the crucial point for a reliable
theory-experiment comparison is the necessity of assessing the real
behavior of the mirrors used in the experiments.

\section{Non specular scattering}

We will now present a more general formalism to calculate the Casimir force
and energy which takes into account non-specular reflection by the plates.
Non-specular reflection is of course the generic reflection process on any
mirror while specular reflection is an idealization.

In order to introduce the more general formula, let us first rewrite
expression (\ref{Eim}) of the Casimir energy between two flat plates
as a sum over modes labeled by the $\xi$ and $m\equiv \bk,p$
\begin{eqnarray}
&&E_\sp  =\hbar\int\limits_{0}^{\infty}\frac{\dd\xi}{2\pi}
\mathrm{Tr} \Delta_\bk^p \left[ i\xi \right] \nonumber\\
&&\Delta_\bk^p \left[ i\xi \right] = \ln\left(
1-r_{1}r_{2}e^{-2\kappa L}\right) \label{Esp}
\end{eqnarray}
This can be interpreted as the energy stored inside the cavity during the scattering process.
It is expressed in terms of the phase shifts $\Delta_\bk^p \left[ i\xi \right]$ acquired
by the field modes upon scattering on the cavity.
These phase shifts are deduced from the $S$-matrix of the cavity \cite{Jaekel91} in such
a manner that the Casimir energy is simply equal to the logarithm of
the determinant of the $S-$matrix.
Using the techniques of quantum field theory \cite{Plunien86}, this can also be written
as the trace of matrix, here diagonal, defined on these modes
\begin{equation}
E_\sp  = \hbar\int\limits_{0}^{\infty}\frac{\dd\xi}{2\pi}\sum_{m}
\left\langle m\right\vert \ln
(1-r_{1}r_{2}e^{-2\kappa_{i}L})\left\vert m\right\rangle
\label{Espec}
\end{equation}
Here, $r_{1}$ and $r_{2}$ are diagonal matrices which contain as
their diagonal elements the specular reflection amplitudes, as they
are seen from fields inside the cavity,
\begin{align}
&  \left\langle m\right\vert r_{i}\left\vert m^{\prime}\right\rangle
\equiv\delta_{m,m^{\prime}}~r_{i}\left(  \xi,m\right)  \nonumber\\
&  \delta_{m,m^{\prime}}\equiv4\pi^{2}\delta\left(\bk-\bk^\prime\right)
\delta_{p,p^{\prime}}
\end{align}
while $\kappa$ is a matrix diagonal over the same modes
\begin{equation}
\left\langle m\right\vert \kappa\left\vert m^{\prime}\right\rangle
\equiv\sqrt{\bk_{m}^{2}+\xi^{2}}\delta_{m,m^{\prime}}
\end{equation}

It is now easy to write down a more general formula of the Casimir energy
for the case of stationary but non-specular scattering
\begin{equation}
E_{\nsp}=\hbar\int\limits_{0}^{\infty}\frac{\dd\xi}{2\pi
}\mathrm{Tr}\ln\left(  1-\mathcal{R}_{1}e^{-\kappa L}\mathcal{R}%
_{2}e^{-\kappa L}\right) \label{Enonspec}
\end{equation}
The two matrices $\mathcal{R}_{1}$ and $\mathcal{R}_{2}$ are no
longer diagonal on plane waves since they describe non specular
reflection on the two mirrors. The propagation factors remain
diagonal on plane waves. Note that the matrices appearing in
(\ref{Enonspec}) no longer commute with each other. As a
consequence, the two propagation matrices in (\ref{Enonspec}) can be
moved through circular permutations in the product but not adjoined to each other.

Formula (\ref{Enonspec}) has already been used to evaluate the
effect of roughness \cite{MaiaEPL05,MaiaPRA05} or corrugation
\cite{MaiaPRL06} of the mirrors on the Casimir force. To this aim,
it was expanded at second order in the profiles of the mirrors, with
the optical response of the bulk metals described by the plasma
model. The non specular reflection amplitudes were then deduced from
techniques developed for treating rough plates
\cite{Agarwal77,Greffet88}. The condition of validity of this
expansion is that the roughness or corrugation amplitude is the
smallest of length scales involved in the problem. In this regime,
it was possible to investigate various domains for the roughness or
corrugation wavelength and thus to investigate the effect of
roughness or corrugation outside as well as inside the range of
validity of the Proximity Force Approximation.

We may again emphasize at this point that the formula (\ref{Enonspec}) has a wider range
of validity than used in those applications.
It can in principle describe mirrors with nanostructured surfaces corresponding
to large amplitudes which cannot be treated as a small perturbation.
It can as well deal with more complicated optical responses which are described neither by a plasma nor by a Drude model.
As was extensively discussed above for the case of specular reflection, the formula
(\ref{Enonspec}) remains valid for arbitrary mirrors, the only problem being to
obtain the precise form of the reflection matrices to be inserted into it.

\subsection{Influence of surface roughness}

Let us now recall how the non-specular scattering formula (\ref{Enonspec})
can be used to calculate the effect of roughness on the Casimir force.
Taking this effect into account simultaneously with that of finite conductivity
is essential, because both of them are important at short distances.
In order to analyze the roughness effect between two metallic plates, we will
describe the optical properties of the mirrors by the plasma model.
The values for the plasma wavelength, the mirror separation and the roughness
correlation length will be arbitrary with respect to each other, the roughness amplitude
remaining the smallest length scale for perturbation theory to hold.
We will review some simple analytical expressions for several limiting cases, as well as
numerical results allowing one for a reliable calculation of the roughness correction
in real experiments \cite{MaiaEPL05,MaiaPRA05}.

\begin{figure}[ptb]
\begin{center} \includegraphics[width=5cm]{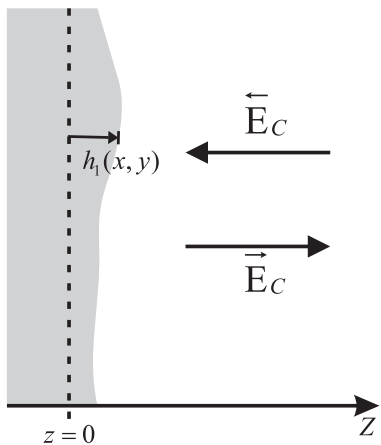}
\caption{Magnified detail of the internal surface of  mirror M1.}
\label{fig7}
\end{center}
\end{figure}

In a plane-plane geometry, the surface profiles are defined by the functions
$h_i(x,y)$ $(i=1,2)$ giving the local heights with respect to the mean separation $L$
along the $z$ direction as shown in Fig.\ref{fig7}.
These functions are defined so that they have zero averages.
We consider the case of stochastic roughness characterized by spectra
\begin{equation}
\sigma_{ij}(\bk) = \int\dd^{2}\br
\,e^{-i\bk\cdot\br}\langle h_i(\br)
\,h_j(\mathbf{0})\rangle \;,\quad i=1,2
\end{equation}
We suppose the surface $A$ of the plates to contain many correlation areas, which allows
us to take ensemble or surface averages interchangeably.
The two plates are considered to be made of the same metal and the crossed
correlation between their profiles is neglected ($\sigma_{12}(\bk)=0$).

We obtain the following variation of the Casimir energy $E_{PP}$ up to second order
in the perturbations $h_i$~\cite{GenetEPL03}
\begin{eqnarray}
\label{main}
&&\delta E _{\rm PP}  =\int\frac{\dd^2\bk}{4\pi^{2}} \GR(\bk)  \sigma(\bk) \\
&&\sigma(\bk)=\sigma_{11}(\bk)+\sigma_{22}(\bk) \nonumber
\end{eqnarray}
With our assumptions, the spectrum $\sigma(\bk)$ fully characterizes the roughness of the two plates.
The correlation length $\ell_C$ is defined as the inverse of its width.
The response function $\GR(\bk)$ then describes the spectral sensitivity to roughness of the Casimir effect.
Symmetry requires that it only depends on $k=|\bk|$.
The dependance of $\GR$ on $k$ reflects that not only the roughness amplitude
but also its spectrum plays a role in diffraction on rough surfaces~\cite{Agarwal77,Greffet88}.
The formula (\ref{main}) has been obtained for the energy in the plane-plane configuration
but it also determines the force correction $\delta F _{\PS}$ in the plane-sphere configuration
since the PFA is still used for describing the weak curvature of the sphere (see below).

We now focus our attention on the validity of PFA for treating the effect of roughness
and notice that this validity only holds at the limit of smooth surface profiles $k\to 0$.
In fact, the following identity is obeyed by our result \cite{MaiaEPL05},
for arbitrary values of $L$ and $\lambda_\P$,
\begin{equation}
\label{G0}
\GR\left(k \to 0\right)=\frac{E_{\PP}''(L)}{2}
\end{equation}
where the derivative is taken with respect to the plate separation $L$.
If we now suppose that the roughness spectrum $\sigma(k)$ is
included inside the PFA sector where $\GR\left(k\right) \simeq \GR\left(0\right)$,
$\GR$ may be replaced by $\GR(0)$ and factored out of the
integral (\ref{main}) thus leading to the PFA expression~\cite{GenetEPL03}
\begin{eqnarray}
\label{eqPFA}
&&\delta E _{\PP}  = \frac{E''_{\PP}(L)}{2} a^2 \\
&&a^2 = \int\frac{\dd^2\bk}{4\pi^2} \sigma(\bk)
\equiv \langle h_1^2 + h_2^2 \rangle \nonumber
\end{eqnarray}
In this PFA limit, the correction depends only on the variance $a^2$ of the roughness
profiles, that is also the integral of the roughness spectrum.

\begin{figure}[ptb]
\begin{center} \includegraphics[width=9cm]{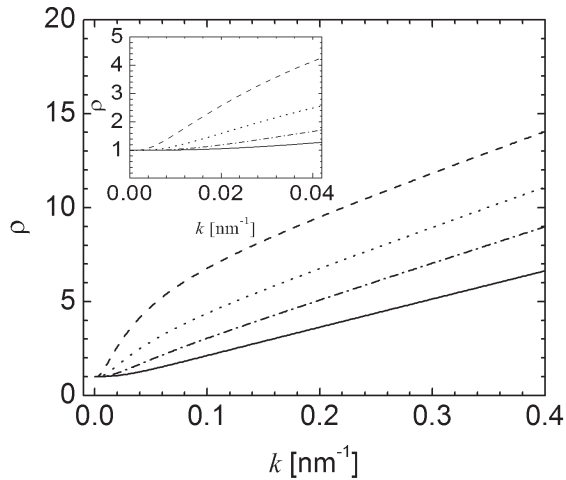}
\caption{Variation of $\rR$ versus $k$ for $L$=50, 100, 200, 400 nm
(from bottom to top curve).} \label{fig3}
\end{center}
\end{figure}

In the general case in contrast, the sensitivity to roughness depends on the
wavevector $k$. This key point is emphasized by introducing a new function $\rR(k)$
which measures the deviation from the PFA~\cite{GenetEPL03}
\begin{equation}
\rR(k) = \frac{\GR(k)}{\GR(0)}.
\end{equation}
This function is plotted on Fig.~\ref{fig3} for several values of $L$.
As for all numerical examples considered below, we take $\lambda_\P=137$nm which
corresponds to gold covered plates.

The ratio $\rR(k)$ is almost everywhere larger than unity, which means that the PFA systematically
\textit{underestimates} the roughness correction.
The inlet shows $\rR(k)$ for small values of $k$ where the PFA is a good approximation.
To give a number illustrating the deviation from the PFA, we find $\rR\simeq 1.6$
for $L=200$nm and $k = 0.02 {\rm nm}^{-1}$,
which means that the exact correction is $60\%$ larger than the PFA result
for this intermediate separation and a typical roughness wavelength
$2\pi/k \simeq 300$nm.

\begin{figure}[ptb]
\begin{center} \includegraphics[width=9cm]{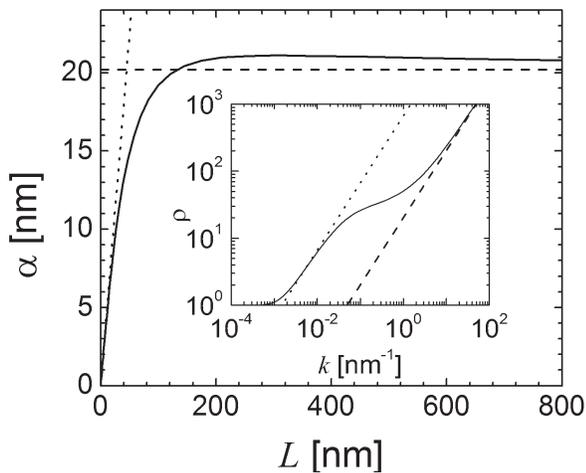}
\caption{Variation of the coefficient $\aR$ versus $L$. The
analytical result for $k^{-1} \ll L \ll \lambda_\P$ is shown as the
dotted line and for $k^{-1} \ll \lambda_\P \ll L$ as the dashed
line. A comparison between this second result (dashed straight line)
and the exact $\rR(k)$ (solid line) is shown in the inlet for $L=2
\mu$m. The analytical result $\rR=L k/3$ predicted by the model of
perfect reflectors (dotted line) is valid only in the intermediate
range $ \lambda_\P\ll k^{-1} \ll L$.} \label{fig8}
\end{center}
\end{figure}

Fig.~\ref{fig3} indicates that $\rR(k)$ grows linearly for large values of $k$.
This is a general prediction of our full calculations \cite{MaiaEPL05},
for arbitrary values of $L$ and $\lambda_\P$,
\begin{equation}
\rR(k)= \aR\, k \quad {\rm for}\quad k \gg {2\pi\over\lambda_\P}\;,\; {1\over L}
\end{equation}
The dimensionless parameter $\aR/L$ depends on $2\pi L/\lambda_P$ only,
and its expression is given by equation (8) in \cite{MaiaEPL05}.
In Fig.~\ref{fig8}, we plot the coefficient $\aR$ as a function of $L$
with $\lambda_\P=137$nm.
In the limit of short distances, we recover the expression which was drawn in
\cite{GenetEPL03} from older calculations \cite{Maradudin80} (after a correction
by a global factor 2)
\begin{equation} \label{kgde-plasmon}
\aR = 0.4492 L \quad {\rm for} \quad k^{-1}\ll L\ll {\lambda_\P\over2\pi}
\end{equation}
In the opposite limit of large distances, the coefficient $\aR$ is found to saturate \cite{MaiaEPL05}
\begin{equation}\label{extreme-k}
\aR = \frac{14}{15} {\lambda_\P\over2\pi}
\quad{\rm for} \quad k^{-1} \ll {\lambda_\P\over2\pi} \ll L
\end{equation}

It is interesting to note that this result differs from the long distance behavior
which was drawn in \cite{GenetEPL03} from the reanalysis of calculations of the effect of
sinusoidal corrugations on perfectly reflecting plates~\cite{EmigPRL01}.
Perfect reflectors indeed correspond to the limiting case where
$\lambda_\P$ rather than $1/k$ is the shortest length scale.
The following result is obtained in this case \cite{MaiaEPL05},
which effectively fits that of ~\cite{EmigPRL01},
\begin{eqnarray}
\label{perfect}
\rR&=&\frac{1}{3}L\,k \quad{\rm for} \quad \lambda_P \ll k^{-1} \ll L
\end{eqnarray}
The long-distance behavior is thus given by (\ref{extreme-k})
when $1/k\ll \lambda_P\ll L$ but by (\ref{perfect}) when $\lambda_P\ll 1/k\ll L$.
The cross-over between these two regimes is shown in the inlet of Fig.~\ref{fig8}, where
we plot $\rR$ as a function of $k$ for $L=2 \mu$m.
The failure of the perfect reflection model for $1/k\ll \lambda_\P$ has been given
an interpretation in \cite{MaiaEPL05}: it results from the fact that not only the
incoming field mode but also the outgoing one have to see the mirror
as perfectly reflecting for formula (\ref{perfect}) to be valid.

These numerical results can be used to assess the accuracy of the
PFA applied to the problem of roughness. PFA is indeed recovered at
the limit of very smooth surface profiles and the deviation from PFA
given by our results as soon as the roughness wavevector goes out of
this limit. The mirrors used in a given experiment have a specific
roughness spectrum which can, and in our opinion must be, measured
when the experiments are performed. The integral (\ref{main}) then
leads to a reliable prediction for the roughness correction, as soon
as the spectral sensitivity $\GR(k)$ and the real spectrum
$\sigma(k)$ are inserted into it.

\subsection{Lateral Casimir force component}

The spectral sensitivity $\GR(k)$ involved in the calculation of the
roughness correction can be considered as a further prediction of
Quantum ElectroDynamics, besides the more commonly studied mean
Casimir force, so that the comparison of its theoretical expectation
with experiments is an interesting prospect. But this comparison can
hardly rely on the roughness correction (\ref{main}) which remains
in any case a small variation of the longitudinal Casimir effect. A
more stringent test can be performed by studying the lateral
component of the Casimir force which arises between corrugated
surfaces. This lateral Casimir force would indeed vanish in the
absence of surface corrugation so that the expression of the
spectral sensitivity will thus appear directly as a factor in front
of the lateral Casimir force. For reasons which will become clear
below, the spectral sensitivity involved in the calculation of
corrugation effect is a different function $\GC(k)$.

Nice experiments have shown the lateral Casimir force to be measurable at separations of a
few hundred nanometers~\cite{ChenPRL02}, that is of the same order of magnitude as the
plasma wavelength $\lambda_\P$. It follows that these experiments can neither be analyzed
by assuming the mirrors to be perfect reflectors~\cite{EmigEPL03}, nor by using
the opposite limit of plasmon interaction~\cite{Maradudin80}.
It is no more possible to use the PFA if we want to be able to treat arbitrary values
of the ratio of the corrugation wavelength $\lambda_\C$ to the interplate distance $L$.
This is why we emphasize the results drawn from the non-specular scattering formula
(\ref{Enonspec}) which can be used for calculating the lateral Casimir force for arbitrary
relative values of $\lambda_\P$, $\lambda_\C$ and $L$.
The only drawback of this calculation is that it is restricted to small enough corrugation
amplitudes, since the latter have to remain the smallest length scale for perturbation
theory to hold. But the lateral force is known to be experimentally accessible in this regime.
Again we model the optical response of the metallic plates by the plasma model.

\begin{figure}[ptb]
\begin{center} \includegraphics[width=5cm]{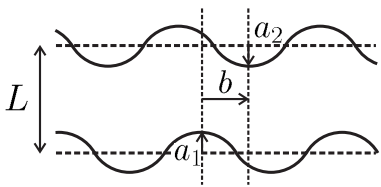}
\caption{Surface profiles considered for the lateral component of
the Casimir force. Both surfaces have a sinusoidal corrugation with
$a_1$ and $a_2$ being the corrugation amplitudes, $b$ the mismatch
between the two sinusoidal functions.} \label{sinus}
\end{center}
\end{figure}

The surface profiles of the corrugated plates are defined by two functions $h_i(\br)$,
with $\br=(x,y)$ is the lateral position along the surfaces of the plates while
$i=1,2$ labels the two plates.
As in experiments~\cite{ChenPRL02}, we consider the simple case of uniaxial sinusoidal
corrugations imprinted on the two plates (see Fig.\ref{sinus}) along the same direction,
say the $y$ direction, and with the same wavevector $k \equiv 2\pi /\lambda_\C$
\begin{eqnarray}
h_{1} =a_1\cos \left(kx\right) \quad,\quad h_{2} =a_2\cos \left(
k(x+b) \right) && \label{profiles}
\end{eqnarray}
Both profiles $h_{1}$ and $h_{2}$ have zero spatial averages and
they are counted as positive when they correspond to local length
decreases below the mean value $L$.

For the purpose of the calculation of the lateral Casimir force, the
non-specular reflection matrix ${\cal R}_j$ have to be developed up
to the first order in the deviations $h_j$ from flatness of the two
plates. They are thus written as the sum of a zero-th order
contribution identifying to the specular reflection amplitude and of
a first-order contribution proportional to the Fourier component at
wavevector $(\bk-\bk^\prime)$ of the surface profiles, this Fourier
component being able to induce a scattering of the field modes from
the wavevector $\bk$ to $\bk'$ ~\cite{MaiaPRL06}. The correction of
the Casimir energy $\delta E_\PP$ induced by the corrugations arises
at second order in the corrugations, with crossed terms of the form
$a_1\,a_2$ which have the ability to induce lateral forces. In other
words, the corrugation sensitivity function $\GC(k)$ obtained below
depends on the crossed correlation between the profiles of the two
plates, in contrast to the function $\GR(k)$ calculated above for
describing the roughness spectral sensitivity. The latter were
depending on terms quadratic in $h_1$ or $h_2,$ and their evaluation
required that second order non specular scattering be properly taken
into account. Here, first order non specular amplitudes evaluated on
both plates are sufficient.

The result of the calculation is read as a second-order correction
induced by corrugations
\begin{eqnarray}
\label{F(k)3}
&&\delta E_\PP=2\int \frac{\dd^{2}\bk}{(2\pi)^2A}{\GC}(\bk)H_{1}(\bk)H_{2}(-\bk)
\end{eqnarray}
with the function $\GC(\bk)$ given by equation~(3) in \cite{MaiaPRL06}.
For isotropic media, symmetry requires ${\GC}(\bk)$ to depend only on the
modulus of the wavevector $k=|\bk|$.
We may also assume for simplicity that the two plates are made of the same
metallic medium.
The energy correction thus depends on the lateral mismatch $b$ between the corrugations
of the two plates, which is the cause for the lateral force to arise.
Replacing the ill-defined $(2\pi)^2 \delta ^{(2)}(0)$ by the area $A$ of the plates,
we derive from (\ref{F(k)3})
\begin{equation}
\delta E_{\PP}= a_1 a_2 \cos (kb) \GC(k)
\label{Epp1/2}
\end{equation}

Once again, the result of the PFA is recovered from equation
(\ref{Epp1/2}) as the limiting case $k\rightarrow0$, that is also
for long corrugation wavelengths. This corresponds to nearly plane
surfaces where the Casimir energy can be obtained from the energy
$E_\PP$ calculated between perfectly plane plates by averaging the
`local' distance ${\cal L}=L-h_1-h_2$ over the surface of the
plates. Expanding at second order in the corrugation amplitudes and
disregarding squared terms in $a_1^2$ and $a_2^2$ because they
cannot produce a lateral dependence, we thus recover expression
(\ref{Epp1/2}) with $\GC(k)$ replaced, for small values of $k$ or
equivalently large values of $\lambda_\C$, by $\GC(0)$ given by
(compare with (\ref{G0}))
\begin{equation}
{\GC}(k \to 0) = {E_\PP^{\prime\prime}\over2}
\label{PFA3}
\end{equation}
This property is ensured, for any specific model of the material
medium, by the fact that $\GC$ is given by the specular limit of non
specular reflection amplitudes~\cite{MaiaPRL06} for $k\rightarrow0$.
For arbitrary values of $k$, the deviation from PFA is then
described by the ratio
\begin{equation}
\rC(k)=\frac{\GC(k)}{\GC(0)}
\label{PFA4}
\end{equation}

\begin{figure}[ptb]
\begin{center} \includegraphics[width=8cm]{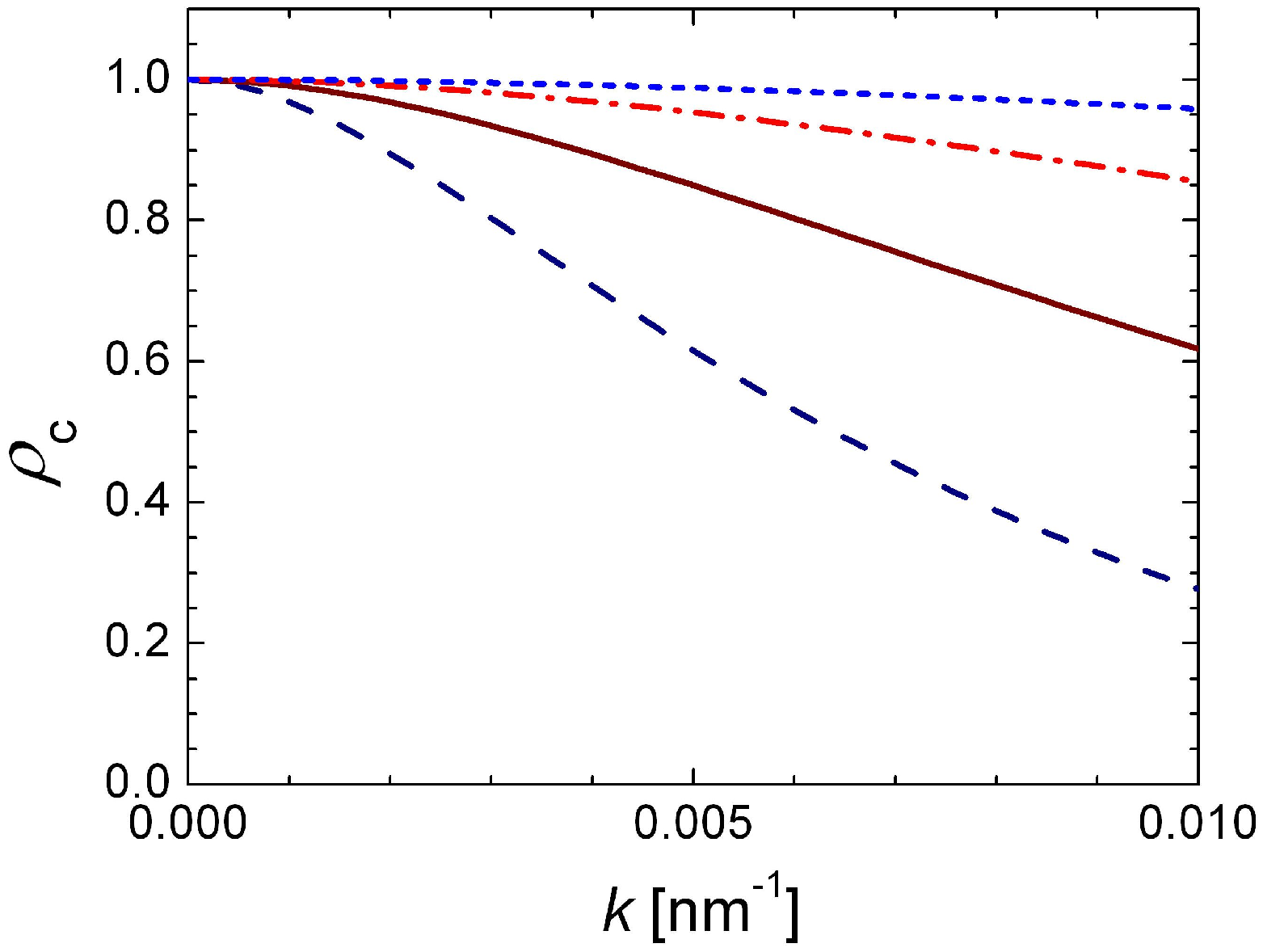}
\caption{ Variation of $\rC$ versus $k$ with $\lambda_\P=137$nm and
for $L=$ 50nm (dotted line), 100nm (dash-dotted line), 200nm (solid
line) and 400nm (dashed line).} \label{ratio}
\end{center}
\end{figure}

In the following, we discuss explicit expressions of this ratio
$\rC$ given by its general expression (eq.~(3) in \cite{MaiaPRL06}).
For the numerical examples, we take $\lambda_\P=137$nm,
corresponding to gold covered plates. The result $\rC$ is plotted on
Fig.~\ref{ratio} as a function of $k$, for different values of the
distance $L$.

For example for a distance $L=50$nm, the Proximity Force Approximation is correct in the range $k\le 0.01 {\rm nm}^{-1}$
({\it i.e.} $\lambda_\C\ge 628$nm) covered by the plot in Fig.~\ref{ratio}.
However, for typical separations of 100nm or larger, $\rC$ drops significantly below its PFA value of unity. A more detailed discussion can be found in \cite{MaiaPRL06}.

\begin{figure}[ptb]
\begin{center} \includegraphics[width=8cm]{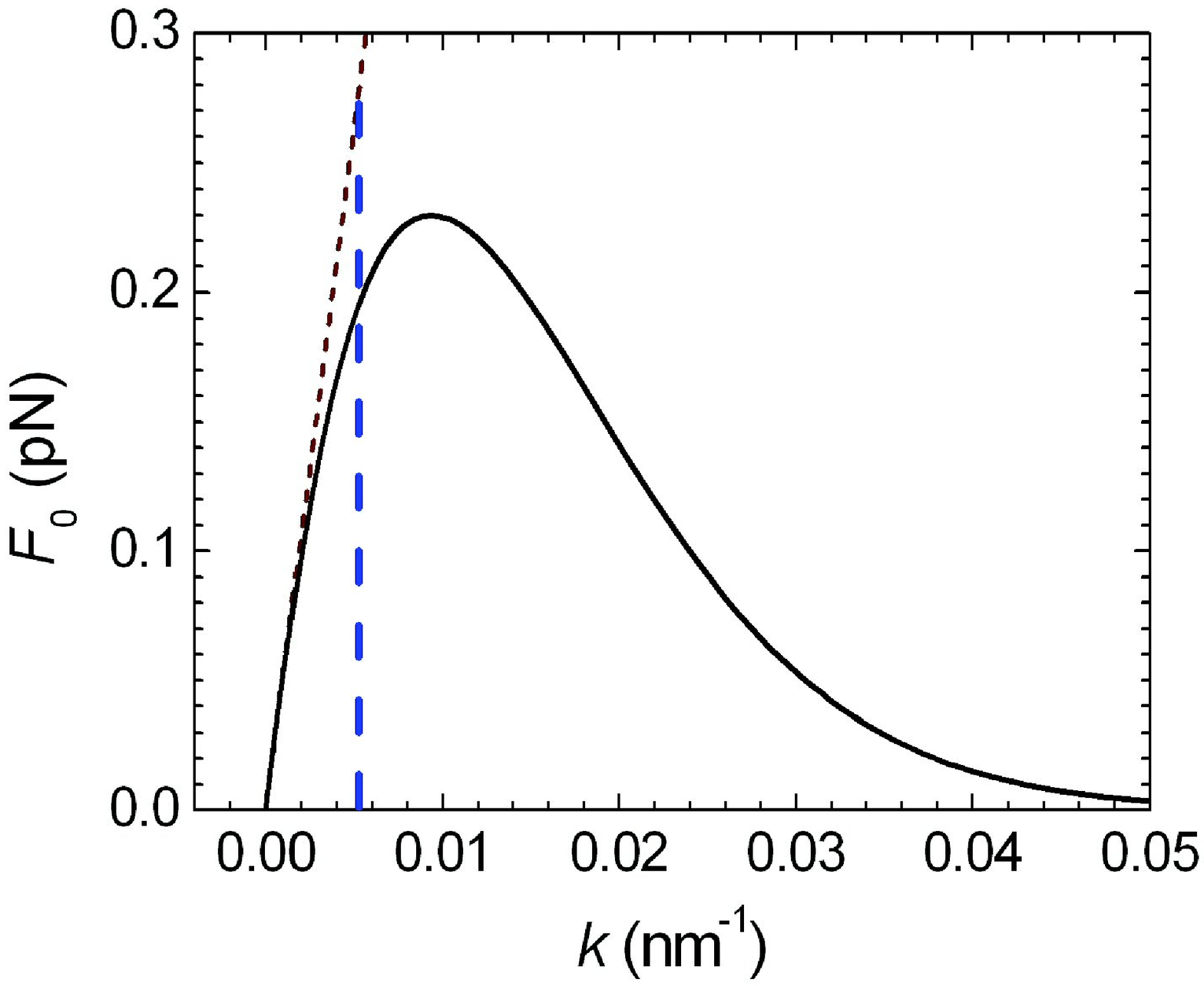} \caption{Lateral
force amplitude for the plane-sphere setup, as a function of $k,$
with figures taken from~\cite{ChenPRL02}. The experimental value
$k=0.0052 {\rm nm}^{-1}$ is indicated by the vertical dashed line. }
\label{vsk}
\end{center}
\end{figure}

For still larger values of $kL$, the functions $\GC(k)$ and
$\rC(k)$ decay exponentially to zero.
If we also assume that $k\lambda_\P\gg 1$, we find
$\GC(k)=\aC \,k\,\exp (-kL) $ where the parameter $\aC $
now depends on $\lambda_\P$ and $L$ only. This is in striking
contrast with the behavior of the response function for stochastic
roughness, which {\it grows} linearly with $k$ for large $k$ due to
the contribution of  the second-order reflection coefficients \cite{MaiaPRA05}.
These coefficients do not contribute to the second-order lateral effect,
which is related to two first-order non-specular reflections at different plates,
separated by a one-way propagation with a modified momentum of the
order of $k$. The resulting propagation factor is, in the large-$k$
limit,  $\exp(-\kappa L)\approx\exp(-kL)$, thus explaining the exponential behavior.

\subsection{Comparison to experiments in a plane-sphere configuration}

In order to compare the theoretical expression of the lateral Casimir
force to experiments, we have to consider the plane-sphere (PS)
geometry~\cite{ChenPRL02} rather than the plane-plane (PP) one.
As $R\gg L$, we use the PFA to connect the two geometries.
Any interplay between curvature and corrugation is avoided provided
that $RL\gg \lambda_\C^2$. These two conditions are
met in the experiment reported in~\cite{ChenPRL02},
where $R=100 \mu$m, $\lambda_\C=1.2 \mu$m and $L\sim 200$nm.

We thus obtain the
energy correction $\delta E_\PS$ between the sphere and a plane at a
distance of closest approach $L$ as an integral of the energy
correction $\delta E_\PP$ in the PP geometry
\begin{equation}
\delta E_\PS(L,b) = \int_{\infty}^{L} \frac{2\pi R\dd L'}{A}
\delta E_\PP(L',b) \label{EPSsmooth}
\end{equation}
Then the lateral force is deduced by varying the energy correction (\ref{EPSsmooth})
with respect to the lateral mismatch $b$ between the two corrugations.
Simple manipulations then lead to the lateral Casimir force in the PS
geometry
\begin{equation}
F^\lat_\PS = \frac{2\pi a_1 a_2}{A} k R \sin(kb) \int_{\infty}^{L}
\dd L' \GC(k,L')
\end{equation}
The force attains a maximal amplitude for $\sin(kb)=\pm1$, which is
easily evaluated in the PFA regime $k\rightarrow0$ where ${\GC}(k)$
does not depend on $k$, so that $F^\lat_\PS$ scales as $k.$
As $k$ increases, the amplitude increases at a slower rate
and then starts to decrease due to the exponential decay of ${\GC}(k)$.
For a given value of the separation $L$, the lateral
force reaches an optimum for a corrugation wavelength such that $kL$
is of order of unity, which generalizes the result obtained for
perfect reflectors in~\cite{EmigPRL01}. In Fig.~\ref{vsk}, we plot the
force $F^\lat_\PS$ (for $\sin(kb)=1$) as a function of $k$, with
figures taken from the experiment of Ref.~\cite{ChenPRL02}. We also use
the values $a_1=59$nm and $a_2=8$nm of the amplitudes for measuring
the force as in~\cite{ChenPRL02}, reminding however that our calculations
are valid in the perturbative limit $a_1,a_2\rightarrow0$.

The plot clearly shows the linear growth for small $k$ as well as
the exponential decay for large $k$. The maximum force is at
$k=0.009{\rm nm}^{-1}$ so that $kL\simeq2$. The experimental value
$k=0.0052 {\rm nm}^{-1}$ is indicated by the dashed line in
Fig.~\ref{vsk}, and the force obtained as 0.20pN, well below the PFA
result, indicated by the straight line and corresponding to a force
of 0.28pN.

Such a variation in the lateral Casimir force should in principle be
measurable in an experiment. This could lead to the first
unambiguous evidence of the limited validity of the Proximity Force
Approximation, that is also to the first observation of a non
trivial effect of geometry on the Casimir force.

\section{Conclusion}

In this review we have described the theory of the Casimir effect using
the techniques of scattering theory.
We have recalled how this formalism allows ones to take into account
the real conditions under which Casimir force measurements are performed.

In particular, the finite conductivity effect can be treated in a
very precise manner, which is a necessity for a reliable
theory-experiment comparison. There however remain inaccuracies in
this comparison if the reflection amplitudes are drawn from optical
models, because of the intrinsic dispersion of optical properties of
samples fabricated by different techniques. We have emphasized that
these inaccuracies could be circumvented by measuring these
reflection amplitudes rather than modeling them.

We have then presented the scattering formulation of the Casimir force
at non zero temperature. This formulation clears out the doubt on the
expression of the force while again requiring to have at one's disposal
reflection amplitudes representing the real properties of the mirrors
used in the experiments.
Let us at this point emphasize that the effect of temperature has not been
unambiguously proven in experiments, and that its observation is one of
the most urgent challenges of experimental research in the domain.
An interesting possibility would be to perform accurate measurements of the
force at distances larger than a few microns, for example by using
torsional balances \cite{LambrechtCQG05}.

In the second part of the paper, we have presented a more general
scattering formalism which takes into account non-specular
reflection. We have also discussed the application of this formalism
for the calculation of the roughness correction to the longitudinal
Casimir force as well as of the lateral component of the Casimir
force arising between corrugated surfaces. We have argued that the
spectral sensitivity functions appearing in these expressions have
to be considered as a new prediction of Quantum ElectroDynamics,
which differ from the more commonly studied mean Casimir force as
soon as one goes out of the domain of validity of the PFA. This new
test seems to be experimentally feasible and constitutes another
challenge of great interest to be faced in the near future.

PAMN thanks R. Rodrigues for discussions and CNPq and Instituto do
Milenio de Informa\c{c}\~ao Quantica for partial financial support.
AL and SR acknowledge fruitful discussions with M.T. Jaekel and C.
Genet. AL acknowledges partial financial support by the European
contract STRP 12142 NANOCASE.

\end{document}